\documentstyle{article} \topmargin=-0.4in \oddsidemargin=-0.2in
\newfont{\set}{msbm10} \newfont{\Set}{msbm6}

\newcommand{\R}{$\mbox{\set R}$} \newcommand{\C}{$\mbox{\set C}$}
\newcommand{\T}{$\mbox{\set R}^{2}$} \newcommand{\V}{$\mbox{\set C}^{d}$}
\newcommand{\W}{$\mbox{\set C}^{2}$}

\newcommand{\be}{\begin{equation}} \newcommand{\ee}{\end{equation}}

\begin{document}

\begin{flushright} DTP/94-39\\ October 1994 \end{flushright} \vspace{4mm}
 \begin{center} { \Large \bf Integrable boundary conditions for classical
 sine-Gordon theory} \end{center} \vspace{3mm} \centerline{A.  MacIntyre }
 \centerline{Department of Mathematical Sciences,} \centerline{University of
 Durham, Durham DH1 3LE, U.K.} \vspace{4mm}

\begin{abstract} The possible boundary conditions consistent with the
integrability of the classical sine-Gordon equation are studied.  A boundary
value problem on the half-line $x\leq 0$ with local boundary condition at
the origin is considered.  The most general form of this boundary condition
is found such that the problem be integrable.  For the resulting system an
infinite number of involutive integrals of motion exist.  These integrals
are calculated and one is identified as the Hamiltonian.  The results found
agree with some recent work of Ghoshal and Zamolodchikov.

\end{abstract}

\section{Introduction}

The problem of factorizable scattering on a half-line necessarily involves
the so-called ``reflection equation'' \cite{chered,fring}.  This equation
supplements the Yang-Baxter relation for the factorization of three particle
scattering and represents the factorization of two particle scattering in
the presence of a one-dimensional `wall' at $x=0$.  Recently there has been
increased interest in this problem, and solutions have been suggested for
sine-Gordon, affine Toda, and $O(N)$ sigma models \cite{zam,Sasi,sas,ghosh}.

 Equations of ``reflection'' type are also used within the framework of the
quantum inverse scattering method \cite{izkor} to prove the integrability of
open quantum spin chains, i.e. chains without periodic boundary conditions
\cite{sklyan,mez,deveg}.  The classical limit of these results was developed
in a paper by Sklyanin \cite{sklcla}, where it was found that they
generalize the Hamiltonian, ($r$-matrix), approach to inverse scattering
theory \cite{tak-fad}, to the case of nonlinear partial differential
equations with local boundary conditions.  The adaptation of these results
to inverse scattering theory on the half line $x\leq 0$ with a local
boundary condition at $x=0$ \cite{bib-tar,tar} saw the problem continued
evenly to the whole line with a `point-spin' impurity at the orgin.  These
results are equivalent to the approach in \cite{Khab1,Khab2} where
symmetrical reductions of B{\"{a}}cklund transformations were considered.
The analysis of these problems on a finite interval is to be found in
\cite{it1,it2}.

Ghoshal and Zamolodchikov \cite{zam}, conjectured both the classical and
quantum integrability of the sine-Gordon theory on the half line with a
boundary term at the origin.  A classical action is, \be
S=\int^{\infty}_{-\infty}\, dt\,\int^{\infty}_{-\infty}\, dx \left[
\Theta(-x)\left(
\frac{1}{2}(\partial_{\mu}\phi(x,t))^2-\frac{m^2}{\beta^2}(1-\cos
\beta\phi(x,t))\right) +\delta(x)\,V(\phi(x,t))\right] , \ee the
minimisation of which leads to the boundary value problem, \begin{eqnarray}
\partial^{\mu}\partial_{\mu}\phi(x,t)+\frac{m^2}{\beta}\sin\beta\phi(x,t)
&=&0\,\,\,\,\,\,\,x\leq 0,\nonumber\\
\frac{\partial\phi(x,t)}{\partial x}+\frac{\partial V(\phi(x,t))}{\partial
\phi}&=&0\,\,\,\,\,\,\,x=0,\nonumber\\ \lim_{x\rightarrow -\infty}
\left(\phi(x,t)\,({\rm{mod}}\,2\pi/\beta)\right)&=&0,\label{eq:sigbc}
\end{eqnarray} with the boundary value at $-\infty$ being attained in the
sense of Schwartz, i.e.  $\phi=0\,({\rm{mod}}\,2\pi/\beta)$ and all its
derivatives decay faster than any inverse power of $x$ as
$x\rightarrow-\infty$.

By demanding that a modification of the first non-trivial integral of motion
of the bulk theory remain conserved on including the boundary it was
found \cite{zam} that in general, \be
V(\phi(x,t))=\frac{Am}{\beta}\cos\left(
\frac{\beta(\phi(x,t)-\phi_{0})}{2}\right),\label{eq:zbc}\ee where $A$ and
$\phi_{0}$ are arbitrary real constants.  In \cite{zam} the existence of an
infinite number of integrals of motion for the system
(\ref{eq:sigbc}),(\ref{eq:zbc}) was assumed and the associated quantum field
theory was studied.

In this paper, the question of integrability of the classical system
(\ref{eq:sigbc}) is addressed.  The results of Sklyanin are used to prove
that an infinite number of integrals of motion
do exist, but only for certain
$V(\phi)$.  It is found that (\ref{eq:zbc}) is the most general boundary
term compatible with the integrability of (\ref{eq:sigbc}).  The results
thus agree with Ghoshal and Zamolodchikov. Further analysis of the
system (\ref{eq:sigbc}),(\ref{eq:zbc}) can be found in \cite{sal}.

In section \ref{sec-perbc} the auxiliary linear problem of inverse
scattering theory is explained.  The case of periodic boundary conditions is
considered first, followed by the calculation of the integrals of motion for
sine-Gordon.  Next follows Sklyanin's extension to other types of boundary
value problem.

Finally, in section \ref{sec-nonper} the formalism developed is applied to
the system (\ref{eq:sigbc}), and the requirement on the form of the boundary
term $V(\phi)$ is deduced.  The integrals of the resulting boundary value
problem are studied and the Hamiltonian is identified.

The notation of \cite{tak-fad} is used throughout the following sections.

\section{The auxiliary linear problem of inverse scattering theory}
\label{sec-perbc} \subsection{Periodic boundary conditions} The
zero-curvature approach to inverse scattering \cite{tak-fad} relies on the
existence of a pair of linear partial differential equations for a $d\times
d$ matrix whose elements are maps from \T$\times$\C\, to \C.  The
solution of such a system subject to a certain initial condition defines the
transition matrix $T(x,y,t,\alpha)$ uniquely.  Suppressing the argument of
$T(x,y,t,\alpha)$, and understanding that $x\geq y$ throughout, the pair of
equations and the initial condition are, \begin{eqnarray}\frac{\partial
T}{\partial x}&=&U(x,t,\alpha)\,\, T\label{eq:zc1}\\ \frac{\partial
T}{\partial t}&=&V(x,t,\alpha)\,\, T-T\,\, V(y,t,\alpha)\label{eq:zc2}\\
T(y,y,t,\alpha)&=&I.\label{eq:ic}\end{eqnarray}
Here, $U(x,t,\alpha)$,$V(x,t,\alpha)$
are $d\times d $ matrices whose elements are
functions of a complex valued field $\Phi(x,t)$, its derivatives, and a
spectral parameter $\alpha\in\,$\C.

The formal solution to (\ref{eq:zc1}),(\ref{eq:ic}) is given by, \be
T(x,y,t,\alpha)={\cal{P}}\exp\left\{\int^{x}_{y}U(x^{\prime},t,\alpha)
\,\,d\,x^{\prime}\right\},\label{eq:po}\ee
where $\cal{P}$ denotes path ordering of noncommuting factors,
\cite{tak-fad}.

Compatibility of (\ref{eq:zc1}) and (\ref{eq:zc2}) gives the zero curvature
condition, \be \frac{\partial U}{\partial t}-\frac{\partial V}{\partial x}
+\left[ U,V\right]=0,\,\,\,\,\,\,\,\forall\,\,\alpha\in {\mbox{\set C}}
\label{eq:zec}\ee and this
should imply the desired nonlinear evolution equation for $\Phi$.

For the sine-Gordon equation, $U$ and $V$ are $2\times 2$ matrices and can be
written in terms of the standard Pauli matrices $\sigma_k,\,\,k=1,2,3$ and a
real valued field $\phi$ as, \begin{eqnarray}
U(x,t,\alpha)&=&\frac{\beta}{4i}\frac{\partial \phi}{\partial
t}\sigma_{3}+\frac{m}{2i}\cosh\alpha\,\sin\left(\frac{\beta\phi}{2}\right)
\sigma_{1}+\frac{m}{2i}\sinh\alpha\,\cos\left(\frac{\beta\phi}{2}\right)
\sigma_{2}\label{eq:zpot1}\\
V(x,t,\alpha)&=&\frac{\beta}{4i}\frac{\partial \phi}{\partial
x}\sigma_{3}+\frac{m}{2i}\sinh\alpha\,\sin\left(\frac{\beta\phi}{2}\right)
\sigma_{1}+\frac{m}{2i}\cosh\alpha\,\cos\left(\frac{\beta\phi}{2}\right)
\sigma_{2}.\label{eq:zpot2}\end{eqnarray}

The imposition of periodic boundary conditions on $\phi(x,t)$ with period
$2L$ ensures periodicity of $V(x,t,\alpha)$ also, and (\ref{eq:zc2})
becomes, \be \frac{\partial T(L,-L,t,\alpha)}{\partial t}=\left[
V(L,t,\alpha), T(L,-L,t,\alpha)\right].  \label{eq:tl}\ee Taking the trace
of (\ref{eq:tl}) and using the chain rule for differentiation gives, \be
\frac{\partial\,\, \ln \left({\rm{tr}}\,T(L,-L,t,\alpha)\right)}{\partial
t}=0,\,\,\,\,\,\,\,\forall \alpha, \label{eq:zccq}\ee which can be expanded
as a Laurent series in $\alpha$ to give an infinite number of integrals of
motion, related to the field $\phi$ by (\ref{eq:po}).  The existence of such
an infinite set is crucial for the integrability of the original evolution
equation, in particular for the construction of action-angle variables for
the theory.

In the limit $L\rightarrow \infty,$ eq.(\ref{eq:tl}) simplifies as $\phi$ and
its derivatives vanish, and the evolution of the matrix elements of $T$, the
transition coefficients, have a simple form.  A detailed treatment of the
auxiliary linear problem (\ref{eq:zc1}), and its inversion, the Riemann
problem, can be found in \cite{tak-fad}.

The alternative formulation to the one outlined above is the so-called
Hamiltonian, or $r$ matrix, approach.  The problem is now regarded as an
infinite dimensional Hamiltonian system with phase space $\cal{M}$.  Points
in this space are determined by pairs of complex $2L$ spatially periodic
functions at fixed time $(\Phi(x,t),\Pi(x,t))$,\,\newline $\Pi$ being the
conjugate momentum to $\Phi$.  The `algebra of observables' $\cal{A}$ on
$\cal{M}$ is defined over \C\ and comprises functionals
$A:{\cal{M}}\rightarrow$\C, bounded everywhere and having smooth functional
derivatives that are $2L$ periodic functions of $x$.  The Poisson bracket on
this algebra is defined at constant time slices by,
\be \left\{ A,B \right\}:=
\int^{L}_{-L}\left(
\frac{\delta A[\Phi(u,t),\Pi(u,t)]}{\delta
\Pi(z,t)}\frac{\delta B[\Phi(v,t),\Pi(v,t)]}{\delta \Phi(z,t)} -\frac{\delta
A[\Phi(u,t),\Pi(u,t)]}{\delta \Phi(z,t)}\frac{\delta
B[\Phi(v,t),\Pi(v,t)]}{\delta
\Pi(z,t)}\right)\,dz,\label{eq:pb}\ee where $A,B\in\cal{A}$ and
$u,v,z$ now lie inside the fundamental domain $[-L,L]$.  This Poisson
bracket can be generalised to act on $d \times d $ matrices, $C,D$ whose
elements are in $\cal{A}$, by replacing $A,B$ in (\ref{eq:pb}) by
$C^{(1)},D^{(2)}$, respectively, where $C^{(1)}:=C\otimes I, D^{(2)}:= I
\otimes D$, and $I$ is the $d\times d$ identity matrix.  As a result of this
notation, $\{ C^{(1)},D^{(2)}\}$ is a $d^2\times d^2$ matrix whose elements
are the Poisson brackets of all possible combinations of matrix elements of
$C,D$.

The Hamiltonian formulation of the problem relies on the existence of an $r$
matrix.  This is a function from \C\ to ${\rm{End}}\,$(\V $\otimes$ \V).  It
is defined in terms of the transition matrix, $T(x,y,t,\alpha)$
(\ref{eq:po}) by the equation, \be
\left\{T^{(1)}(x,y,t,\alpha),T^{(2)}(x,y,t,\gamma)\right\}=\left[
r_{1\,2}(\alpha -\gamma)
,T^{(1)}(x,y,t,\alpha)T^{(2)}(x,y,t,\gamma)\right],\,\,\,\,\,\forall
\alpha,\gamma\in {\mbox{\set C}}.  \label{eq:fpb}\ee
The dependence of $r$ on the difference
$(\alpha-\gamma)$ is not the case for all models, but it is so for the
sine-Gordon equation considered below.  Antisymmetry and the Jacobi identity
for the bracket hold if and only if, \begin{eqnarray}
r_{1\,2}(\alpha-\gamma)+r_{2\,1}(\gamma-\alpha)&=&0,\\ \left[
r_{1\,2}(\alpha-\gamma),r_{1\,3}(\alpha)+r_{2\,3}(\gamma)\right]+\left[
r_{1\,3}(\alpha),r_{2\,3}(\gamma)\right]&=&0,\label{eq:cybe}\end{eqnarray}
where the subscripts denote the nontrivial action of $r$ in \V $\,\otimes$
\V $\,\otimes$ \V.  Eq(\ref{eq:cybe}) is the classical Yang-Baxter equation.

 For sine-Gordon, phase space ${\cal{M}}_{\scriptscriptstyle{SG}}$ is the
subset of $\cal{M}$ given by pairs of {\em{real}} $2L$ periodic functions
$(\phi(x,t),\pi(x,t))$, \be \phi(x+2L,t)\equiv\phi(x,t)\bmod
\frac{2\pi}{\beta},\,\,\,\,\,\,\,\,\pi(x+2L,t)=\pi(x,t),\,\,\,\,\pi(x,t):=
\frac{\partial\phi(x,t)}{\partial t}.\label{eq:sgps}\ee
On this space will be defined the
algebras ${\cal{A}}_{\scriptscriptstyle{SG}}$ and
${\cal{A}}_{\scriptscriptstyle{SG}}^{\mbox{\Set R}}$.  The algebra
${\cal{A}}_{\scriptscriptstyle{SG}}$ is the restriction of $\cal{A}$ to
${\cal{M}}_{\scriptscriptstyle{SG}}$, while
${\cal{A}}_{\scriptscriptstyle{SG}}^{\mbox{\Set R}}$ is the subset of
${\cal{A}}_{\scriptscriptstyle{SG}}$ mapping
${\cal{M}}_{\scriptscriptstyle{SG}}\rightarrow\,$\R, which becomes an
algebra on restricting the scalars of ${\cal{A}}_{\scriptscriptstyle{SG}}$
to \R.  For convenience, the Poisson bracket will be defined on both these
algebras but it is understood that the integrals of motion should be
elements of ${\cal{A}}_{\scriptscriptstyle{SG}}^{\mbox{\Set R}}$.  Also,
$d=2$ and, \be
r_{\scriptscriptstyle{SG}}(\alpha-\gamma)=
\frac{\beta^2}{16\tanh(\alpha-\gamma)}
(I\otimes I-\sigma_{3}\otimes
\sigma_{3})-\frac{\beta^2}{16\sinh(\alpha-\gamma)}(\sigma_{1}\otimes\sigma_{1}
+\sigma_{2}\otimes\sigma_{2}).\ee

Returning to the integrals of motion of a general theory, taking the trace
of (\ref{eq:fpb}) and using the chain rule for functional differentiation
gives, \be\left\{\ln
\left({\rm{tr}}\,T(x,y,t,\alpha)\right),\ln\left({\rm{tr}}
\,T(x,y,t,\gamma)\right)\right\}=0,\,\,\,\,\,\forall\,\alpha,\gamma\in
\mbox{\set C},\,\,\,\,\,x,y\,\,\,{\rm{s.t.}}\,\, -L\leq y\leq x\leq L
 \label{eq:invch}\ee which, on expansion as
Laurent series in $\alpha,\gamma$ about the singularities in $T$, gives an
infinite number of involutive charges, depending on $\Phi$ through
(\ref{eq:po}).  However, so far no mention has been made of the boundary
conditions on $\Phi$.  These have been used in defining the `algebra of
observables' $\cal{A}\,$, and the fundamental domain \newline $-L\leq y\leq
x\leq L$.  For these involutive quantities to have smooth functional
derivatives and so be elements of this algebra requires that $x=L,y=-L$.
Hence, the integrals of motion found using the zero-curvature formulation
are also in involution with respect to the Poisson bracket (\ref{eq:pb}).
This suggests that the model may be completely integrable, along the lines
of the Liouville-Arnold theorem \cite{arn} for finite dimensional
Hamiltonian systems.

Alternatively, there is no need to appeal to the zero-curvature condition at
all.  By requiring the involutive charges to be elements of $\cal{A}$, they
are fixed uniquely. Among these should exist a Hamiltonian $H(L,-L,t)$ such
that, \be \frac{\partial\Phi(x,t)}{\partial
t}=\left\{H(L,-L,t),\Phi(x,t)\right\},\,\,\,\,\frac{\partial\Pi(x,t)}{\partial
t}=\left\{H(L,-L,t),\Pi(x,t)\right\},\,\,\,-L\leq x\leq L,\label{eq:hem}\ee
imply the field equation.  The problem can then be extended to the $x$ axis by
letting $L\rightarrow\infty$ and demanding that $\Phi$ be a Schwartz
function.

For the sine-Gordon equation such a Hamiltonian is found as follows. Fixing
${\rm{Im}}\,\alpha=\pi/2$, (\ref{eq:zpot1}) shows $T(L,-L,t,\alpha)$ to have
essential singularities at ${\rm{Re}}\,\alpha=\pm\infty$.  As a result, it
is expected that there will exist two distinct infinities of involutive
integrals coming from the coefficients of the Laurent expansions about these
two ``points''.  First consider ${\rm{Re}}\,\alpha\rightarrow +\infty$.
Suppressing $t$ and defining $T^{\prime}(L,-L,\lambda)=T(L,-L,\alpha)$ and
$\lambda:=e^\alpha$, $T^{\prime}(x,y,\lambda)$ is rewritten as, \be
T^\prime(x,y,\lambda)=\Omega(x)\tilde{T}(x,y,\lambda)\Omega^{-1}(y),
\label{eq:gt}\ee where $\Omega(x)=\exp(i\beta\phi(x)\sigma_{3}/4)$.  It can
be shown \cite{tak-fad} that for $\lambda\rightarrow\, \pm i\infty$, \be
\tilde{T}(x,y,\lambda)= \left(I+W(x,\lambda)\right)\exp
Z(x,y,\lambda)\left(I+W(y,\lambda)\right)^{-1}+O(|\lambda|^{-\infty}),
\label{eq:ae}\ee
where, \begin{eqnarray}
W(x,\lambda)&=&\sum^{\infty}_{n=0}\frac{W_{n}(x)}{\lambda^n},\label{eq:Wdef}\\
[0.15in] Z(x,y,\lambda)&=&\frac{m\lambda
(x-y)}{4i}\sigma_3+i\sum^{\infty}_{n=1}\frac{Z_{n}(x,y)}{\lambda^n}.
\label{eq:Zmat}
\end{eqnarray} The $W_{n},Z_{n}$ have the form, \be
W_{n}(x)=\left(\begin{array}{cc} 0&-\bar{w}_{n}(x) \\ w_{n}(x)&0 \end{array}
\right),\,\,\, Z_{n}(x,y)=\left(\begin{array}{cc} z_{n}(x,y)&0 \\
0&-\bar{z}_{n}(x,y) \end{array} \right), \label{eq:WZmat}\ee with, \be
w_{0}(x)=i,\ee \begin{eqnarray} w_{n+1}(x)&=&\frac{2}{im}\frac{\partial
w_{n}(x)}{\partial x}-\frac{\beta}{m}\left(\pi(x)
+\frac{\partial \phi(x)}{\partial x}\right)
w_{n}(x)+\frac{i}{2}\sum^{n}_{k=1}w_{k}(x)w_{n+1-k}(x)\nonumber\\
&\,&\,\,-\frac{i}{2}e^{-i\beta\phi(x)}\sum^{n-1}_{k=0}w_{k}(x)w_{n-k-1}(x)-
\frac{i}{2}e^{i\beta\phi(x)}\delta_{n,1},\label{eq:wn}\\
z_{n}(x,y)&=&\frac{im}{4}\int^{x}_{y}(w_{n+1}(x')-
e^{-i\beta\phi(x')}w_{n-1}(x'))d\,x'.
\label{eq:zn}\end{eqnarray}

Now imposing periodic boundary conditions at $\pm L$ simplifies the form of
${\rm{tr}}\,\,T'(L,-L,\lambda)$.  As a result, the first set of integrals
will be the coefficients of $ {\rm{tr}}\, \exp Z(L,-L,\lambda).$ The
unimodularity of $T'(L,-L,\lambda)$ implies that $Z(L,-L,\lambda)$ be
traceless up to $O(|\lambda|^{-\infty})$ terms, and so the $z_{n}(L,-L)$'s
are real and $Z(L,-L,\lambda)$ is proportional to $\sigma_{3}$.  Thus, \be
{\rm{tr}}\,
T'(L,-L,\lambda)=\frac{1}{2}\cos\left(\sum_{n=1}^{\infty}
\frac{z_{n}(L,-L)}{\lambda^{n}}-
\frac{m\lambda L}{2}\right)
+O(|\lambda|^{-\infty}),\,\,\,\, \lambda\rightarrow\pm
i\infty.\label{eq:exptra}\ee Rewriting the cosine in terms of exponentials
and neglecting the one which is $O(|\lambda|^{-\infty})$, the result is,
 \be
\ln \left({\rm{tr}}\,T'(L,-L,\lambda)\right)= i\left[\frac{m\lambda L}{2}-
\frac{z_{1}(L,-L)}{\lambda}-\frac{z_{2}(L,-L)}{\lambda^{2}}
-O(|\lambda|^{-3})\right],
\,\,\,\,\,\,\lambda\rightarrow+i\infty,\label{eq:lntra}\ee
and this series generates the set
of charges $\{z_{n}(L,-L)\}$ defined by (\ref{eq:wn}),(\ref{eq:zn}).  These
charges are purely imaginary, time independent, in involution with each
other, and are elements of ${\cal{A}}_{\scriptscriptstyle{SG}}$.  To
construct elements of ${\cal{A}}_{\scriptscriptstyle{SG}}^{\mbox{\Set R}}$
it suffices to multiply all the charges by $i$.  Note that
(\ref{eq:ae}),(\ref{eq:exptra}) also hold for $\alpha=-\pi/2$ or
$\lambda\rightarrow -i\infty$.  The resulting set of charges in
${\cal{A}}_{\scriptscriptstyle{SG}}$ are complex conjugate to those found
above and lead to the same elements of
${\cal{A}}_{\scriptscriptstyle{SG}}^{\mbox{\Set R}}$ on multiplication by
$-i$.

The second distinct set of integrals comes from the expansion of
${\rm{tr}}\,T'(L,-L,\lambda)$ as $\lambda\rightarrow i0\pm$.  To find this,
use is made of the invariance of (\ref{eq:zpot1}),(\ref{eq:sgps}) under the
replacement $\pi(x,t)\rightarrow\pi(x,t),\,\phi(x,t)\rightarrow
-\phi(x,t),\,e^{\alpha}\rightarrow-e^{-\alpha}$.  As a result \be
\hat{T'}\left(x,y,-\frac{1}{\lambda}\right)=T'(x,y,\lambda),\label{eq:invT}\ee
where $\hat{T'}(x,y,\lambda)$ denotes the transition matrix for
$\hat{\pi}(x,t)=\pi(x,t),\, \hat{\phi}(x,t)=-\phi(x,t)$.  So taking
$\lambda\rightarrow i0\pm$ in (\ref{eq:invT}) amounts to taking
$\mu:=-1/\lambda \rightarrow \pm i\infty$ in $\hat{T'}(L,-L,\mu)$.
Therefore the analysis given above for $\lambda\rightarrow \pm i\infty$ can
be used again on the replacement
$\lambda\rightarrow-1/\lambda,\phi(x,t)\rightarrow-\phi(x,t)$.  The
resulting conserved charges are, \be
\hat{z}_n(L,-L)=\frac{im}{4}\int^{L}_{-L}(\hat{w}_{n+1}(x')-
e^{i\beta\phi(x')}
\hat{w}_{n-1}(x'))d
x',\ee with the $\hat{w}_{n}$'s given by (\ref{eq:wn}) on replacing
$\phi\rightarrow -\phi$.  Repeating the same procedure as above for this
second set shows the $\{\hat{z}_{n}(L,-L)\}$ to be elements of
${\cal{A}}_{\scriptscriptstyle{SG}}^{\mbox{\Set R}}$.

The linear combination
$H(L,-L):=\frac{2m}{\beta^2}(z_{1}(L,-L)+\hat{z}_{1}(L,-L))$ is of the form,
\be
H(L,-L)=\int^{L}_{-L}\left(\frac{1}{2}\pi^2+\frac{1}{2}\left(\frac{\partial
\phi}{\partial
x}\right)^{2}+\frac{m^{2}}{\beta^{2}}(1-\cos\beta\phi)\right)d\,x,\ee and is
identified as the Hamiltonian of the model.  The sine-Gordon equation
results from (\ref{eq:hem}).  Note that all the
$z_{n}(L,-L),\hat{z}_{n}(L,-L)$ are zero for $n$ even.  Also note that
if $\phi$ is an even function of $x$ then the Hamiltonian density is itself
even.  In general the densities
$z_{n}(L,-L)+{\hat{z}}_{n}(L,-L),\,\,\,\,n=1,3,..$ have this property,
(``parity even'').  Finally extending $L\rightarrow\infty$ and demanding
that $\phi$ have Schwartz boundary conditions, gives the theory on the whole
line.

\subsection{Local boundary conditions} \label{subsec-kmat} The analysis
given below was first provided by Sklyanin \cite{sklcla}.  The idea is to
modify the Hamiltonian approach of the last subsection by introducing a new
`algebra of observables' $\cal{B}$, the form of which is not specified.
This is to be contrasted with the precise definition of the algebra
$\cal{A}$ given previously.  By analogy with the transition matrix
$T(x,y,t,\alpha)$, a $d\times d$ matrix with elements in $\cal{B}$ is
introduced and called ${\cal{T}} (x,y,t,\alpha)$.  This matrix is required
to satisfy the equation, cf.(\ref{eq:fpb}), \be
\left\{{\cal{T}}^{(1)}(\alpha),{\cal{T}}^{(2)}(\gamma)\right\}=\left[
r_{1\,2}(\alpha-\gamma),{\cal{T}}^{(1)}(\alpha){\cal{T}}^{(2)}(\gamma)\right]
+{\cal{T}}^{(1)}(\alpha)r_{1\,2}(\alpha+\gamma){\cal{T}}^{(2)}(\gamma)
-{\cal{T}}^{(2)}(\gamma)r_{1\,2}(\alpha+\gamma){\cal{T}}^{(1)}(\alpha),
\label{eq:mpb}\ee
$\forall\,\,\alpha,\gamma,$ and the $x,y,t$ dependence has been suppressed.
The Poisson bracket is
defined in (\ref{eq:pb}) with the exception that the range of the integral
is changed to $[x_{-},x_{+}]$, and it is understood that $x_{-}\leq y\leq
x\leq x_{+}$.  Antisymmetry of this bracket implies, \begin{eqnarray}
r_{1\,2}(\alpha)-r_{2\,1}(\alpha)&=&0\nonumber\\
r_{1\,2}(-\alpha)+r_{1\,2}(\alpha)&=&0.  \end{eqnarray}

 Constant representations of the Poisson bracket algebra $\cal{B}$ must also
satisfy (\ref{eq:mpb}), hence \be \left[
r_{1\,2}(\alpha-\gamma),K^{(1)}_{\pm}(\alpha)K^{(2)}_{\pm}(\gamma)\right]+
K^{(1)}_{\pm}(\alpha)
r_{1\,2}(\alpha+\gamma)K^{(2)}_{\pm}(\gamma)-K^{(2)}_{\pm}(\gamma)
r_{1\,2}(\alpha+\gamma)K^{(1)}_{\pm}
(\alpha)=0,\label{eq:kref}\ee where $K_{\pm}(\alpha)$ are $d\times d$ \C\
numerical matrices of the complex spectral parameter $\alpha$ and possibly
some complex constants.  From such a $K_{-}(\alpha)$, a
${\cal{T}}(x,y,t,\alpha)$ satisfying (\ref{eq:mpb}) can be built according
to, \be {\cal{T}}(x,y,t,\alpha)\equiv
T(x,y,t,\alpha)K_{-}(\alpha)T^{-1}(x,y,t,-\alpha),\label{eq:nrep}\ee where
$T(x,y,t,\alpha)$ is given by (\ref{eq:po}).

Multipying ${\cal{T}}(x,y,t,\alpha)$ on the left by $K_{+}(\alpha)$,
evaluating the bracket of this with itself using (\ref{eq:mpb}), taking the
trace and using the chain rule for functional differentiation gives, \be
\left\{
\ln\left({\rm{tr}}\,K_{+}(\alpha){\cal{T}}(x,y,t,\alpha)\right),
\ln\left({\rm{tr}}\,K_{+}(\gamma)
{\cal{T}}(x,y,t,\gamma)\right)\right\}=0,\,\,\,\,\,\forall\,\alpha,\gamma\in
\mbox{\set C},\,\,\,\,\,x,y\,\,\,{\rm{s.t}}\,\,\, x_{-}
\leq y\leq x\leq x_{+},\label{eq:IIOM}\ee
i.e.  an infinite set of involutive quantities. The
question remains, however, as to whether these can be used to construct
elements of the unspecified algebra $\cal{B}$.  As was the case for the
algebra $\cal{A}$, the requirement of smooth functional derivatives of
elements of $\cal{B}$ imposes $y=x_{-},\,x=x_{+}$.  In principle,
specification of the boundary conditions at $x_{\pm}$ on elements of phase
space should be sufficient to determine $\cal{B}$ uniquely.  In contrast to
the periodic case, however, local boundary conditions of the form
$\left.\partial_{x}\Phi+\partial_{\Phi}V(\Phi)\right|_{x=x_{\pm}}=0$ do not
impose any general form for functional derivatives of elements of $\cal{B}$.
So, given boundary conditions at $x_{\pm}$, the entire form of $\cal{B}$
cannot be deduced on symmetry grounds, there are no constraints on the forms
of the $K_{\pm}(\alpha)$, and a construction analogous to the one leading
from (\ref{eq:invch}) to (\ref{eq:hem}) cannot be performed.

The way around this problem is to appeal to the zero-curvature formulation.
For periodic boundary conditions, (\ref{eq:zccq}) can be regarded as a
consistency check on the form of ${\rm{tr}}\,T(x,y,t,\alpha)$ such that it
be an element of $\cal{A}$.  In this case, such a check just imposes
$y=-L,x=L$ which gives no new information on the structure of the algebra.
If the same point of view is taken for the nonperiodic case, however,
$K_{\pm}(\alpha)$ become constrained by possible forms of boundary condition
at $x_{\pm}$.  To see this, consider, $\partial_{t}
K_{+}(\alpha){\cal{T}}(x_{+},x_{-},t,\alpha)$ using
(\ref{eq:zc2}),(\ref{eq:nrep}).  The trace of this quantity should be zero
if the Hamiltonian and zero-curvature approaches are to be consistent.  This
requirement leads to, \be
K_{\pm}(\alpha)V(x_{\pm},t,\pm\alpha)=V(x_{\pm},t,\mp\alpha)K_{\pm}(\alpha),
\,\,\,\,\,\forall
\alpha\in\mbox{\set C},\label{eq:kvc}\ee which contains nontrivial
information as to the forms of the $K_{\pm}(\alpha)$, for particular
boundary conditions of the theory.

\section{Application to sine-Gordon theory} \label{sec-nonper} In this
section the formalism developed in subsection \ref{subsec-kmat} is applied
to the sine-Gordon equation.  For sine-Gordon $d=2$ and the matrices $U,V,$
appearing in the zero-curvature condition are given by
(\ref{eq:zpot1}),(\ref{eq:zpot2}).  The basis of ${\rm{End}}\,$\W\, is
chosen to be the Pauli matrices and the two dimensional identity
($:=\sigma_{0}$), and $K_{\pm}(\alpha)$ is written in this
basis as $D_{\pm}^{\mu}(\alpha)\sigma_{\mu},\,\,\mu=0,..,4$ .  It should be
recalled that the standard Pauli matrices satisfy
$\sigma_{i}\sigma_{j}=i\epsilon_{ijk} \sigma_{k}+\delta_{ij}\sigma_{0},\,\,
i,j,k=1,..,3$.  Substituting $K_{\pm},V$ into (\ref{eq:kvc}) gives, \be
\left.D_{\pm}^{\mu}(\alpha)\left(\frac{\beta}{4i}\frac{\partial\phi}{\partial
x}\left[\sigma_{\mu},\sigma_{3}\right]
+\frac{m}{2i}\sinh\alpha\sin\left(\frac{\beta\phi}{2}\right)\left
\{\sigma_{\mu},\sigma_{1}\right\}
+\frac{m}{2i}\cosh\alpha
\cos\left(\frac{\beta\phi}{2}\right)\left[\sigma_{\mu},\sigma_{2}\right]
\right)\right|_{x=x_{\pm}}=0.\ee
The anticommutator implies $D_{\pm}^{1}(\alpha)=0$, and so the most general
form this equation can take is, \be \left.  \frac{\partial\phi}{\partial
x}-\frac{2mP_{\pm}(\alpha)}{\beta}\sin\left(\frac{\beta\phi}{2}\right)-
\frac{2mQ_{\pm}(\alpha)}
{\beta}\cos\left(\frac{\beta\phi}{2}\right)\right|_{x=x_{\pm}}=0,
\label{eq:sic}\ee
where, \be P_{\pm}(\alpha)=
\frac{iD_{\pm}^{0}(\alpha)}{D_{\pm}^{2}(\alpha)}\sinh\alpha,\,\,\,\,\,
\,\,Q_{\pm}(\alpha)=
\frac{D_{\pm}^{3}(\alpha)}{D_{\pm}^{2}(\alpha)}\cosh\alpha.\ee
Alternatively, $K_{\pm}(\alpha)=D^{0}_{\pm}\sigma_{0}$ implies, \be
\left.\frac{R_{\pm}(\alpha)m}{2\beta}
\sin\left(\frac{\beta\phi}{2}\right)\right|_{x=x_{\pm}}=0
,\label{eq:scz}\ee
with, \be R_{\pm}(\alpha)=\frac{D^{0}_{\pm}(\alpha)\sinh\alpha}{i}.\ee

 From now on only the boundary at $x_{+}$ will be considered; however
similar arguments would apply to the other boundary.  The subscripts $\pm$ can
therefore be suppressed.  Eq.(\ref{eq:sic}) would be in the form of a
boundary condition that could be imposed on $\phi$ if $P(\alpha),Q(\alpha),$
were {\em{real}} and independent of $\alpha$.  Similarly (\ref{eq:scz})
would be a boundary condition if $R(\alpha)$ were independent of $\alpha$.
Denoting these constants by $P,Q,R$ respectively, the form of $K(\alpha)$ is
constrained even more.  The result is, \be
K(\alpha)=\frac{D^{2}(\alpha)}{i\sinh\alpha}\left(P\sigma_{0}+
i\sinh\alpha\sigma_{2}
+iQ\tanh\alpha\sigma_{3}\right),\label{eq:KTL}\ee
or, \be K(\alpha)=\frac{iR}{\sinh\alpha}\sigma_{0}.\label{eq:KTZ}\ee
It can be verified
that these satisfy (\ref{eq:kref}) so that (\ref{eq:IIOM}) holds with
(\ref{eq:KTL}),(\ref{eq:KTZ}).

 Since the $K(\alpha)$ are only defined up to an overall scale, it will be
convenient to work with, \be\tilde{K}(\alpha):=
\frac{P}{\sinh\alpha}\sigma_{0}+i\sigma_{2}
+\frac{iQ}{\cosh\alpha}\sigma_{3},\ee and \be
\tilde{K}(\alpha):=\sigma_{0}.\ee The normalization is fixed by requiring
$\det \tilde{K}(\alpha)\rightarrow 1$ as $\rm{Re}\,\alpha\rightarrow\pm
\infty$.  So, by demanding that $\phi$ satisfy boundary conditions of the
form, \be\left.  \frac{\beta}{2m}\frac{\partial\phi}{\partial x}-P_{\pm}
\sin \left(\frac{\beta\phi}{2} \right)-Q_{\pm} \cos \left(
\frac{\beta\phi}{2} \right) \right|_{x=x_{\pm}}=0, \label{eq:zerbc}\ee or,
\be\left.
\phi-\frac{2n\pi}{\beta}\right|_{x=x_{\pm}}=0,\,\,\,\,\,\,\,\,\,
n=1,2,\ldots,\label{eq:swb}\ee
the associated $\tilde{K}_{\pm}(\alpha)$ can be found and an infinite number
of integrals in involution can be constructed.  These will be the
coefficients in the expansion of, \be
\ln\left({\rm{tr}}\,\tilde{K}_{+}(\ln(\lambda))T'(x_{+},x_{-},\lambda)
\tilde{K}_{-}
(\ln(\lambda)){T'}^{-1}(x_{+},x_{-},1/\lambda)\right),\label{eq:ddd}\ee
as $\lambda\rightarrow \pm i\infty,\,i0\pm$.  From these charges, elements of
the algebra of observables ${\cal{B}}_{\scriptscriptstyle{SG}}^{\mbox{\Set
R}}$ will be constructed as they were for periodic boundary conditions.

Now, consider the boundary value problem (\ref{eq:sigbc}).  Taking
$x_{-}=-\infty$, (\ref{eq:swb}) becomes the required Schwartz boundary
condition.  With $x_{+}=0$, (\ref{eq:zerbc}) has the form of
(\ref{eq:sigbc}),(\ref{eq:zbc}) after relabelling constants, and the desired
boundary value problem has been fitted into the developed formalism.  Hence
the conclusion that (\ref{eq:sigbc}), (\ref{eq:zbc}) is integrable.  It
remains to construct the algebra of observables and identify a Hamiltonian
for the problem from the infinity of integrals.  This is done below.

First put the problem (\ref{eq:sigbc}),(\ref{eq:zbc}) on the finite interval
$[-L,0]$ and define \begin{eqnarray}& & {\tilde{K}}_{0}(\alpha):=
\frac{P}{\sinh\alpha}\sigma_{0}+i\sigma_{2}+\frac{iQ}{\cosh\alpha}\sigma_{3},
\label{eq:KT}\\
& &\,\,\,\,\,{\tilde{K}}_{-L}(\alpha):=\sigma_{0},\end{eqnarray} so that
(\ref{eq:ddd}) becomes, \be
\ln\left({\rm{tr}}\,\tilde{K}_{0}(\ln(\lambda))T'(0,-L,\lambda){\tilde{K}}_{-L}
T'(-L,0,1/\lambda)\right).\label{eq:expp}\ee
To expand this in an asymptotic series as $\lambda\rightarrow \pm i\infty$,
use is made of (\ref{eq:invT}),(\ref{eq:gt}),(\ref{eq:ae}) giving,
\begin{eqnarray}& &\ln\left({\rm{tr}}\,{\tilde{K}}_{0}
(\ln(\lambda))\Omega(0)(I+W(0,\lambda))\exp
Z(0,-L,\lambda)(I+W(-L,\lambda))^{-1}\Omega^{-1}(-L)\hat{\Omega}(-L)\right.
\nonumber\\
& &\,\,\,\,\,\,\,\,\,\,\,\,\left.\times(I+\hat{W}(-L,-\lambda))\exp\hat{Z}
(-L,0,-\lambda)
(I+\hat{W}(0,-\lambda))^{-1}\hat{\Omega}^{-1}(0)+O(|\lambda|^{-\infty})\right),
\end{eqnarray}
where, $W,Z$ are defined by (\ref{eq:Wdef})-(\ref{eq:zn}) and the hats
indicate the same quantities but with $\phi\rightarrow-\phi$.  Imposing the
Schwartz boundary condition at $-L$ simplifies this further, the result
being, \begin{eqnarray} &
&\ln\left({\rm{tr}}\,{\tilde{K}}_{0}(\ln(\lambda))T'(0,-L,\lambda)
{\tilde{K}}_{-L}(\ln(\lambda))
{T'}^{-1}(0,-L,1/\lambda)\right)\nonumber\\
& &\,\,\,\,\,\,\,\,=\ln\left({\rm{tr}}\,(I+\hat{W}(0,-\lambda))^{-1}
\Omega(0){\tilde{K}}_{0}(\ln(\lambda))\Omega(0)(I+W(0,\lambda))
\exp(Z(0,-L,\lambda)-\hat{Z}
(0,-L,-\lambda))\right)\label{eq:exbc}\end{eqnarray}
mod$(O|\lambda|^{-\infty})$ as $\lambda\rightarrow \pm i\infty$.  From
(\ref{eq:Zmat}), it follows that, \be
Z(0,-L,\lambda)-{\hat{Z}}(0,-L,-\lambda)=i\sum^{\infty}_{n=-1}
\frac{a_{n}(0,-L)}{\lambda^{n}}
\sigma_{3}-\sum_{n=1}^{\infty}\frac{b_{n}(0,-L)}{\lambda^{n}}I,\ee
where
$z_{n}(0,-L)-(-1)^{n}{\hat{z}}_{n}(0,-L)=a_{n}(0,-L)+ib_{n}(0,-L),
\,\,n=1,2,..$,
and $a_{0}(0,-L)=0,\newline a_{-1}(0,-L)=-m\lambda L/2$.  Hence, only the
diagonal terms in
$(I+\hat{W}(0,-\lambda))^{-1}\Omega(0){\tilde{K}}_{0}(\ln(\lambda))\Omega(0)
(I+W(0,\lambda))$
will contribute to the trace.  Defining, \be
{\rm{diag}}\left((I+\hat{W}(0,-\lambda))^{-1}\Omega(0){\tilde{K}}_{0}
(\ln(\lambda))
\Omega(0)(I+W(0,\lambda)\right)=
\sum^{\infty}_{n=0}\left(\frac{ic_{n}(0)}{\lambda^{n}}\sigma_{3}
+\frac{d_{n}(0)}{\lambda^{n}}I\right),\label{eq:diagtr}\ee
where (\ref{eq:KT}),(\ref{eq:exbc}) and the involutions, \begin{eqnarray} &
&\!\!\!\!\!\!\!\!\!\!\!\!\!\!\!\!\!\!\overline{(I+\hat{W}(0,-\bar{\lambda}))
^{-1}\Omega(0)
{\tilde{K}}_{0}(\ln(\bar{\lambda}))\Omega(0)(I+W(0,\bar{\lambda}))}\nonumber\\
& &\,\,\,\,\,\,\,\,\,\,\,\,\,\,\,\,\,\,\,\,\,\,\,\,\,\,
\,\,\,\,\,\,\,\,\,\,\,\,\,\,\,\,\,
\,\,\,\,\,\,\,\,\,\,\,\,\,\,\,\,\,
\,\,\,\,\,\,\,\,\,\,\,\,\,\,\,\,\,\,\,\,\,\,\,\,\,\,\,\,
=\sigma_{2}(I+\hat{W}(0,-\lambda))^{-1}\Omega(0){\tilde{K}}_{0}(\ln(\lambda))
\Omega(0)(I+W(0,\lambda))\sigma_{2},\\ [0.15in] &
&\,\,\overline{\exp(Z(0,-L,\bar{\lambda}))-\hat{Z}(0,-L,-\bar{\lambda})}=
\sigma_{2}\exp(Z(0,-L,\lambda)-\hat{Z}(0,-L,-\lambda))\sigma_{2},
\end{eqnarray} are used to show that the $\{c_{n}(0),d_{n}(0)\}$ are real.
Thus (\ref{eq:exbc}) has the form, \be
\ln\left(2\sum_{n=0}^{\infty}\frac{d_{n}(0)}{\lambda^{n}}
\cos\left(\sum_{m=-1}^{\infty}
\frac{a_{m}(0,-L)}{\lambda^{m}}\right)
-2\sum_{n=0}^{\infty}\frac{c_{n}(0)}{\lambda^{n}}
\sin\left(\sum_{m=-1}^{\infty}
\frac{a_{m}(0,-L)}{\lambda^{m}}\right)\right)
-\sum_{k=1}^{\infty}\frac{b_{k}(0,-L)}{\lambda^{k}},\label{eq:reexp}\ee
mod$(O(|\lambda|^{-\infty})$ as $\lambda\rightarrow \pm i\infty$, the
Laurent expansion of which has real coefficients.  Now consider the
expansion of this as $\lambda\rightarrow +i\infty$.  Rewriting the
$\sin,\,\cos$ in terms of exponentials and discarding the ones which are
$O(|\lambda|^{-\infty})$, (\ref{eq:reexp}) becomes, \be
i\sum_{m=-1}^{\infty}\frac{a_{m}(0,-L)}{\lambda^{m}}-
\sum_{k=1}^{\infty}\frac{b_{k}(0,-L)}
{\lambda^{k}}+\ln\left(\sum_{n=0}^{\infty}\frac{d_{n}(0)
+ic_{n}(0)}{\lambda^{n}}\right),
\label{eq:loge}\ee
as $\lambda\rightarrow +i\infty$. From
(\ref{eq:diagtr}),(\ref{eq:KT}),(\ref{eq:Wdef}),(\ref{eq:WZmat}) one finds
$c_{0}=1,\,d_{0}=0,$ so that the logarithm can be expanded as a power
series.  The coefficient of (\ref{eq:loge}) at $O(1/\lambda)$ is
$i(a_{1}-d_{1})-(b_{1}-c_{1})$.  This is, in principle, a complex valued
functional and is {\em{defined}} to be an element of
${\cal{B}}_{\scriptscriptstyle{SG}}$.  Indeed, this will be the case to all
orders.  It remains to construct the integrals of motion which are elements
of ${\cal{B}}_{\scriptscriptstyle{SG}}^{\mbox{\Set R}}$.  To do this, use is
made of the $\lambda\rightarrow-i\infty$ expansion of (\ref{eq:reexp}) which
has coefficients at each order in $\lambda$ that are complex conjugate to
those in the $\lambda\rightarrow +i\infty$ expansion.  Hence, not only are
the complex integrals in involution with one another, but also the complex
conjugated ones too.  (This is an obvious consequence of the real
coordinates on ${\cal{M}}_{\scriptscriptstyle{SG}}$).  As a result, the real
and imaginary part of each integral is separately in involution with the
real and imaginary parts of all the other integrals.  Therefore once a
Hamiltonian for the system is identified as being the real or imaginary part
at some order in $\lambda$, then the real or imaginary parts of the others
can be chosen at will.  Thus, by multiplying terms by $i$ an infinite number
of elements of ${\cal{B}}_{\scriptscriptstyle{SG}}^{\mbox{\Set R}}$ in
involution, can be constructed.  At $O(1/\lambda)$ in (\ref{eq:loge}), the
first element of ${\cal{B}}_{\scriptscriptstyle{SG}}$ is, \be
\frac{-i\beta^2}{2m}\left(\int^{0}_{-L}\left[\frac{1}{2}\pi^2+
\frac{1}{2}\left(\frac{\partial\phi}{\partial
x}\right)^{2}+\frac{m^{2}}{\beta^{2}}(1-\cos\beta\phi)\right]d\,x
+\frac{4Pm}{\beta^2}\cos\frac{\beta\phi(0)}{2}-\frac{4Qm}{\beta^2}
\sin\frac{\beta\phi(0)}{2}\right).\ee
Multiplication of this by $2mi/\beta^{2}$ {\em{defines}} an element of
${\cal{B}}_{\scriptscriptstyle{SG}}^{\mbox{\Set R}}$ and is identified as
the Hamiltonian of the system (\ref{eq:sigbc}),(\ref{eq:zbc}) defined over
the interval $[-L,0]$.  By imposing Schwartz boundary conditions at $-L$ the
integral can be extended to the whole half-axis.

The $O(1/\lambda^2)$ term is found to be real and equal to $4(P^{2}+Q^{2})$,
i.e.  a trivial integral of motion.  The coefficient at $O(1/\lambda^{3})$
is purely imaginary and is made up of the integral over $[-L,0]$ of the next
``parity even'' density plus a boundary term.  It is not clear as to whether
the series continues to alternate between real and imaginary terms.
However, it will be the case that the integrals at
$O(1/\lambda^{2n}),\,n=1,2,..$,  will depend purely on
the boundary at $x=0$.

Finally it remains to return to (\ref{eq:expp}) to calculate its expansion
as $\lambda\rightarrow i0\pm$.  Replacing $\lambda$ by $-1/\mu$ in
(\ref{eq:expp}) and using (\ref{eq:invT}) one finds that the analysis for
$\lambda\rightarrow\pm i\infty$ will hold on replacing $\phi\rightarrow
-\phi,\,Q\rightarrow -Q$.  So all the integrals previously found will form
another infinite set on imposition of this transformation.  However the
first three are invariant under this, and hence it is likely to be the case
in general.

\section{Conclusions} The system (\ref{eq:sigbc}),(\ref{eq:zbc}) conjectured
to be integrable by Ghoshal and Zamolodchikov has been fitted in to a
formalism developed by Sklyanin for integrable equations with local boundary
conditions.  It has been shown that an infinite number of integrals of
motion, in involution, do exist.  These charges fall into two classes:
those consisting of integrals of ``even parity'' densities plus a boundary
term, and those depending purely on the field and its derivatives at the
boundary.

This analysis fits into the scheme developed by Khabibullin on symmetrical
reductions of B{\"{a}}cklund transformations, but it remains to continue the
problem evenly to the whole axis with a ``point-spin'' at the origin along
the lines of \cite{bib-tar,tar}.  The analysis of the solutions of such a
system is an interesting open problem.

\vspace{0.65 cm} \noindent {\bf Acknowledgments:}  I would like to thank
Peter Bowcock, Ed Corrigan, Patrick Dorey, and Martin Speight for useful
discussions.  I also acknowledge the financial support of the U.K.
Engineering and Physical Sciences Research Council.


\newpage \begin{thebibliography}{xx}

\bibitem{chered} I.V.  Cherednik, ``Factorizing particles on a half line and
root systems'', {\sl{Theor.  Math.  Phys.}}\, {\bf{61(1)}} (1984) 977-983.
\bibitem{fring} A.  Fring and R.  K\"{o}berle, ``Factorized scattering in
the presence of reflecting boundaries'', {\sl{Nucl.  Phys.}}\,{\bf{B421}}
(1994) 159.  \bibitem{zam} S.  Ghoshal and A.B.Zamolodchikov, ``Boundary
state and boundary S matrix in two-dimensional integrable field theory'',
{\sl{Int.  J.  Mod.  Phys.}}\, {\bf{A9}} (1994) 3841.
\bibitem{Sasi} R.  Sasaki, ``Reflection bootstrap equations for Toda
field theory'', in {\sl{Interface between physics and mathematics}}
\ \ eds. W. Nahm, Jian-Min Shen, (World Scientific 1994),
 hep-th/9311027.
 \bibitem{sas} E. Corrigan, P.E.  Dorey, R.H.  Rietdijk, R. Sasaki,
 ``Affine Toda field theory
on a half line'', {\sl{Phys. Lett.}}\,{\bf{B333}} (1994) 83,
``Apects of affine Toda field theory on a half line'', DTP 94/29
 hep-th/9407148.
\bibitem{ghosh} S.  Ghoshal, ``Boundary S matrix of the O(N) nonlinear sigma
model'', {\sl{Phys. Lett.}}\,{\bf{B334}} (1994) 363.
\bibitem{izkor} V.E.  Korepin,
N.M.  Bogoliubov, A.G.  Izergin, {\sl{Quantum inverse scattering method and
correlation functions}} (C.U.P.  Cambridge, U.K.  1992).  \bibitem{sklyan}
E.K.  Sklyanin, ``Boundary conditions for integrable quantum systems'',
{\sl{J.  Phys.}}\,{\bf{A21}} (1988) 2375-2389.  \bibitem{mez} L.  Mezincescu
and R.  Nepomechie, ``Fusion procedure for open chains'', {\sl{J.
Phys.}}\,{\bf{A25}}(1992) 2533-2543.  \bibitem{deveg} H.J.  de Vega and A.
Gonz\'{a}lez Ruiz, ``Boundary K-matrices for the XYZ, XXZ and XXX spin
chains, LPTHE-PAR 93-29 hep-th/9306089.
 \bibitem{sklcla} E.K.  Sklyanin, ``Boundary
conditions for integrable equations'', {\sl{Funct.  Anal.
Appl.}}\,{\bf{21}} (1987) 164.  \bibitem{tak-fad} L.D.  Faddeev and L.A.
Takhtajan, {\sl{Hamiltonian methods in the theory of solitons}}
(Springer-Verlag, Berlin, Germany, 1987).  \bibitem{bib-tar} P.N.  Bibikov
and V.O.  Tarasov, ``Boundary value problem for the nonlinear
Schr\"{o}dinger equation'', {\sl{Theor.  Math.  Phys.}}\,{\bf{79(3)}} (1989)
570-579.  \bibitem{tar} V.O.  Tarasov, ``The integrable initial-boundary
value problem on a semiline:  nonlinear Schr\"{o}dinger and sine-Gordon
equations'', {\sl{Inverse Problems}}\,{\bf{7}} (1991) 435-449.
\bibitem{Khab1} I.T.  Khabibullin, ``Integrable initial-boundary value
problems'', {\sl{Theor.  Math.  Phys.}}\,{\bf{86(1)}} (1991) 28-36.
\bibitem{Khab2} I.T.  Khabibullin, ``The Backlund transformation and
integrable initial-boundary value problems'', {\sl{Math.
Notes}}\,{\bf{49(3-4)}} (1991) 418-423.  \bibitem{it1} R.F.  Bikbaev and
A.R.  Its, ``Algebrogeometric solutions of a boundary value problem for the
nonlinear Schr\"{o}dinger equation'',{\sl{Math.  Notes}}\,{\bf{45(5-6)}}
(1989) 349-354.  \bibitem{it2} R.F.  Bikbaev and A.R.  Its,
``Algebrogeometric solutions of the nonlinear boundary value problem on a
segment for the sine-Gordon equation'', {\sl{Math.  Notes}}\,{\bf{52(3-4)}}
(1992) 1005-1011.
\bibitem{sal} H. Saleur, S. Skorik, N.P. Warner, ``The boundary sine-Gordon
theory: classical and semiclassical analysis'', USC-94-013 hep-th/9408004.
\bibitem{arn} V.I.  Arnold, {\sl{Mathematical methods of
classical mechanics}} (Springer-Verlag, New York, 1978).

\end{thebibliography}
 \end{document}